\documentclass[twocolumn]{aastex63}

\usepackage{xspace, comment, amsmath}

\newcommand{\name}{SN~2011fe\xspace}
\newcommand{\tmax}{$t_{\rm{max}}$\xspace}
\newcommand{\eplus}{$e^+$\xspace}
\newcommand{\Mch}{$M_{\rm{Ch}}$\xspace}
\newcommand{\MWD}{$M_{\rm{WD}}$\xspace}

\newcommand{\Ti}{$^{44}$Ti\xspace}
\newcommand{\Sc}{$^{44}$Sc\xspace}

\newcommand{\decaychainfour}{${^{44}\rm{Ti}\rightarrow^{44}\rm{Sc}\rightarrow^{44}\rm{Ca}}$\xspace}

\newcommand{\CaII}{[\ion{Ca}{2}]\xspace}

\newcommand{\fFe}[1]{[\ion{Fe}{#1}]\xspace}

\newcommand{\iNi}[1]{$^{#1}$Ni\xspace}
\newcommand{\iCo}[1]{$^{#1}$Co\xspace}
\newcommand{\iFe}[1]{$^{#1}$Fe\xspace}
\newcommand{\iMn}[1]{$^{#1}$Mn\xspace}

\newcommand{\rhoc}{$\rho_c$\xspace}

\newcommand{\M}[1]{$M_{#1}$\xspace}
\newcommand{\fopt}{$f_{\rm{opt}+\rm{NIR}}$\xspace}
\newcommand{\rhot}{$\langle\rho\rangle t^3$\xspace}
\newcommand{\Lstar}{$L_\star$\xspace}

\newcommand{\snia}{SN Ia\xspace}
\newcommand{\sneia}{SNe Ia\xspace}

\newcommand{\kms}{km~s$^{-1}$\xspace}
\newcommand{\um}{\ensuremath{\mu}m\xspace}
\newcommand{\AAA}{\AA\xspace}

\newcommand{\gamrays}{$\gamma$-rays\xspace}

\newcommand{\hst}{HST\xspace}
\newcommand{\filtB}{$F438W$\xspace}
\newcommand{\filtV}{$F555W$\xspace}
\newcommand{\filtR}{$F600LP$\xspace}

\received{June 1, 2019}
\revised{January 10, 2019}
\accepted{\today}
\submitjournal{ApJ}

\shorttitle{SN~2011fe}
\shortauthors{M. A. Tucker et al.}
\graphicspath{{./}{figures/}}

\begin{document}

\title{The Whisper of a Whimper of a Bang: 2400 Days of the \\Type Ia SN 2011fe Reveals the Decay of $^{55}$Fe}

\correspondingauthor{Michael Tucker}
\email{tuckerma95@gmail.com}

\author[0000-0002-2471-8442]{M. A. Tucker}
\altaffiliation{DOE CSGF Fellow}
\affiliation{Institute for Astronomy,
University of Hawai`i at Manoa,
2680 Woodlawn Dr., Honolulu, HI, USA}

\author{B. J. Shappee}
\affiliation{Institute for Astronomy,
University of Hawai`i at Manoa,
2680 Woodlawn Dr., Honolulu, HI, USA}

\author{C. S. Kochanek}
\affiliation{Department of Astronomy, The Ohio State University, 140 West 18th Avenue, Columbus, OH 43210, USA}
\affiliation{Center for Cosmology and AstroParticle Physics, The Ohio State University, 191 W. Woodruff Ave., Columbus, OH 43210, USA}

\author{K. Z. Stanek}
\affiliation{Department of Astronomy, The Ohio State University, 140 West 18th Avenue, Columbus, OH 43210, USA}
\affiliation{Center for Cosmology and AstroParticle Physics, The Ohio State University, 191 W. Woodruff Ave., Columbus, OH 43210, USA}

\author{C. Ashall}
\affiliation{Institute for Astronomy,
University of Hawai`i at Manoa,
2680 Woodlawn Dr., Honolulu, HI, USA}

\author[0000-0002-5259-2314]{G. S. Anand}
\affiliation{Space Telescope Science Institute, 3700 San Martin Drive, Baltimore, MD 21218, USA}

\author{P. Garnavich}
\affiliation{Physics Department, University of Notre Dame, Notre Dame, IN 46556, USA}



\begin{abstract}

We analyze new multi-filter \textit{Hubble Space Telescope} (HST) photometry of the normal Type Ia supernova (SN Ia) 2011fe out to $\approx 2400$~days after maximum light, the latest observations to date of a SN Ia. We model the pseudo-bolometric light curve with a simple radioactive decay model and find energy input from both \iCo{57} and \iFe{55} are needed to power the late-time luminosity. This is the first detection of \iFe{55} in a SN Ia. We consider potential sources of contamination such as a surviving companion star or delaying the deposition timescale for \iCo{56} positrons but these scenarios are ultimately disfavored. The relative isotopic abundances place direct constraints on the burning conditions experienced by the white dwarf (WD). Additionally, we place a conservative upper limit of $< 10^{-3}~M_\odot$ on the synthesized mass of \Ti.  Only 2 classes of explosion models are currently consistent with all observations of \name: 1) the delayed detonation of a low-\rhoc, near-\Mch ($1.2-1.3~M_\odot$) WD, or 2) a sub-\Mch ($1.0-1.1~M_\odot$) WD experiencing a thin-shell double detonation. 

\end{abstract}



\section{Introduction}\label{sec:intro}

Type Ia supernovae (SNe Ia) are used as extragalactic distance indicators \citep[e.g., ][]{tully2016} and are key cosmological probes \citep[e.g., ][]{riess1998, perlmutter1999}. Our understanding of \sneia has advanced considerably in the past decade due to the proliferation of sky surveys \citep[e.g., ][]{shappee2013, chambers2016, tonry2018, bellm2019} and advances in numerical models \citep[e.g., ][]{ropke2018, nouri2019}, but key questions remain unanswered. While it is generally accepted that \sneia originate from the thermonuclear disruptions of carbon/oxygen (C/O) white dwarfs (WDs; \citealp{hoyle1960}), how and why the WD explodes as a \snia is unclear. The progenitor system must include a nearby companion star to destabilize the WD but the type of companion star remains debated \citep[see ][for recent reviews]{maoz2014, jha2019}. Additionally, how the flame ignites and propagates in the WD core, referred to as the the explosion mechanism, has not yet reached a consensus \citep[e.g., ][]{hillebrandt2013}, further complicating efforts to determine the progenitor systems of \sneia. 

For many years the standard theory for \snia progenitor systems invoked a non-degenerate star transferring mass onto the WD through Roche Lobe overflow (RLOF), referred to as the `single degenerate' (SD) scenario \citep[e.g., ][]{whelan1973, nomoto1982, yoon2003}. This was driven by the overall homogeneity of \sneia near maximum light \citep[e.g., ][]{folatelli2010}, as mass-transfer from a nearby donor star would naturally ignite WDs as they approach the Chandrasekhar mass ($M_{\rm{Ch}} \approx 1.44~M_\odot$; \citealp{chandrasekhar1931}). Additionally, WDs in mass-transfer binary systems are seen in the Milky Way \citep[e.g., ][]{pala2020}, readily providing both a progenitor system and explosion mechanism. 

However, the SD theory places a non-degenerate star in close proximity to the WD at the time of explosion. Interactions between the donor star and the ejecta are expected to produce several observational signatures \citep[e.g., ][]{wheeler1975} including irregularities in the rising light curve \citep{kasen2010}, supersoft X-ray emission during accretion onto the WD \citep[e.g., ][]{langer2000, schwarz2011}, and radio and/or X-ray emission produced by the high-velocity ejecta interacting with nearby circumstellar material (CSM). Additionally, the high-velocity ejecta will deposit energy into the companion star \citep{marietta2000, pan2012a, boehner2017} increasing its luminosity for several thousand years after explosion \citep[e.g., ][]{pan2012b, shappee2013b} and stripping/ablating $0.1-0.5~M_\odot$ of material from the star \citep{mattila2005, botyanszki2018, dessart2020}.

Searches for SD progenitor systems in normal \sneia have repeatedly returned null results, including no irregularities in the rising light curves \citep{hayden2010, bianco2011, fausnaugh2021}, stringent upper limits on radio \citep[e.g., ][]{panagia2006, chomiuk2016, cendes2020, harris2021} and X-ray \citep[e.g., ][]{margutti2012, russell2012, margutti2014,shappee2018, kilpatrick2018, sand2021} emission indicating a clean circumstellar environment, non-detections of material stripped from the companion star \citep{leonard2007, graham2017, maguire2018, tucker2020}, the lack of high-energy radiation (i.e., supersoft X-ray sources) expected for accreting WDs \citep[e.g., ][]{gilfanov2010, woods2017, woods2018, kuuttila2019}, and ``missing'' surviving companion stars \citep[e.g., ][]{shaefer2012, edwards2012, kerzendorf2018, do2021}. Several studies have claimed detections of a SD progenitor system \citep[e.g., ][]{marion2016, hosseinzadeh2017} but follow-up studies have yet to unambiguously confirm these claims \citep{maguire2016, shappee2018, sand2018, tucker2019}. There is a subclass of \sneia that exhibit signs of interaction with nearby CSM \citep[SNe Ia-CSM; ][]{silverman2013} but these systems are rare and constitute a small fraction of \sneia \citep{graham2019, dubay2021}. Thus, it seems unlikely that most \sneia originate from the SD scenario. 

The alternative is the double degenerate (DD) scenario, where the companion is a degenerate star, typically another WD \citep[e.g., ][]{tutukov1979, iben1984, webbink1984}. However, most double WD binaries do not have a combined mass exceeding \Mch \citep[e.g., ][]{maoz2012} leading to a resurgence of interest in sub-\Mch WDs ($M_{\rm{WD}} \lesssim 1.2~M_\odot$) producing \sneia. Recent observational \citep[e.g., ][]{mazzali2011, scalzo2014, flors2020} and theoretical \citep[e.g., ][]{blondin2017, goldstein2018, townsley2019, polin2019} studies support sub-\Mch producing some fraction of ``normal'' \sneia, but it remains unclear if it is the \textit{dominant} channel for \sneia production. While the Milky Way WD merger rate may be able to reproduce the expected \snia rate \citep[e.g., ][]{maoz2018}, some WD mergers will produce a more massive WD instead of exploding as a \snia \citep[e.g., ][]{giammichele2012, brown2016}. Thus, it remains unclear which parts of the parameter space (i.e., individual WD masses, orbital separation, presence of external bodies) available to double WD binaries lead to successful explosions.

The late-time light curves  of SNe Ia can provide insight into the explosion conditions because they depend on the nucleosynthetic yields \citep[e.g., ][]{seitenzahl2017, graur2018}. The primary energy source for \sneia is the radioactive decay of unstable isotopes synthesized during the explosion \citep{pankey1962, truran1967}. The important decay chains are \citep{colgate1969, seitenzahl2009}

\begin{eqnarray}\label{eq:decaychains}
& &^{56}\mathrm{Ni}  \;\stackrel{t_{1/2} = \; 6.08d}{\hbox to 60pt{\rightarrowfill}} \; ^{56}\mathrm{Co} \; 
\stackrel{t_{1/2} = \; 77.2d}{\hbox to 60pt{\rightarrowfill}} \; ^{56}\mathrm{Fe}, \nonumber\\*
& &^{57}\mathrm{Ni}  \;\stackrel{t_{1/2} = \; 35.60 h}{\hbox to 60pt{\rightarrowfill}}\; ^{57}\mathrm{Co} \;
\stackrel{t_{1/2} = \; 271.79d}{\hbox to 60pt{\rightarrowfill}} \; ^{57}\mathrm{Fe},\\*
& &^{55}\mathrm{Co}  \;\stackrel{t_{1/2} = \; 17.53 h}{\hbox to 60pt{\rightarrowfill}}\; ^{55}\mathrm{Fe} \;
\stackrel{t_{1/2} = \; 999.67 d}{\hbox to 60pt{\rightarrowfill}} \; ^{55}\mathrm{Mn},\nonumber\\
& &^{44}\mathrm{Ti}  \;\stackrel{t_{1/2} = \; 58.9 y}{\hbox to 60pt{\rightarrowfill}}\; ^{44}\mathrm{Sc} \;
\stackrel{t_{1/2} = \; 3.97 h}{\hbox to 60pt{\rightarrowfill}} \; ^{44}\mathrm{Ca}.\nonumber
\end{eqnarray}

\noindent The mass of the synthesized species determines the bolometric evolution and the observed light curve provides insight into the amount of synthesized material \citep[e.g., ][]{arnett1982}. Specifically, the evolution of the late-time light curve is determined by the burning conditions in the WD core. Mass ratios of these unstable isotopes can be derived from late-time photometry but observations at $\gtrsim 2000$~days are necessary to probe the long half-life of \iFe{55}.

\name, discovered a mere $\approx 11$~hours after explosion \citep{nugent2011} by the Palomar Transient Facility \citep[PTF; ][]{law2009}, is the brightest \snia since the advent of modern astronomical detectors. Located at just $d_L \approx 6.5~\rm{Mpc}$ \citep[e.g., ][]{shappee2011,beaton2019}, \name exploded in a region of M101 uncontaminated by intervening dust \citep{patat2013} providing an ideal location for testing \snia progenitor and explosion models. The early detection allowed extensive follow-up observations across the electromagnetic spectrum \citep[e.g., ][]{matheson2012, parrent2012, pereira2013, hsiao2013, johansson2013, tsvetkov2013, munari2013,  mazzali2014, zhang2016} and provided direct constraints on the radius of the exploding star \citep{nugent2011, bloom2012}. Stringent non-detections in radio \citep{chomiuk2012, horesh2012, kundu2017} and X-ray \citep{margutti2012, horesh2012} observations exclude nearby CSM at high significance. Early ultraviolet (UV) photometry did not show any evidence for the ejecta encountering a nearby companion star \citep{brown11fe} and nebular spectra lacked the Balmer emission lines from material ablated off the donor star by the ejecta impact \citep{shappee2013, lundqvist2015, tucker2022}. Pre-explosion imaging excludes the presence of a RG or He donor star \citep{li2011} and disfavor an accreting WD in the $\sim 10^5$~years prior to explosion \citep{graur2014}. Multi-epoch spectropolarimetry reveal consistently-low continuum polarization suggestive of a symmetric ejecta distribution with evidence for minor oblateness \citep{milne2017}. Finally, nebular-phase observations at $\gtrsim 1$~year after maximum light allows a direct view to the inner ejecta and provides unique constraints on the explosion conditions \citep{mcclelland2013,kerzendorf2014, mazzali2015, graham2015, taubenberger2015, kerzendorf2017, dimitriadis2017, shappee2017, friesen2017, tucker2022}. \name is one of the best-studied astronomical objects in the past decade and remains a key benchmark for any \snia theory or model.

In this paper we extend the rich dataset for \name with new HST photometry extending to $\approx 2400$~days after maximum light. Details about the data reduction are provided in \S\ref{sec:data} which we use to construct a pseudo-bolometric light curve in \S\ref{sec:phot}. The viability of various interpretations for the photometric evolution are assessed including inferred isotopic yields, the efficiency of leptonic confinement, and the presence of a surviving companion. With these new observations and analyses, we consider various proposed progenitor systems and explosion mechanisms in \S\ref{sec:discuss}. We adopt the photometric parameters derived by \citet{zhang2016}, in particular the time of maximum $B$-band flux $t_{\rm{max}} = \rm{MJD}~55813.98\pm0.03$. All phases are given in the rest-frame and the rise-time of $\approx 18$~days is taken from \citet{pereira2013}. We correct all observations for Milky Way extinction $E(B-V)_{\rm{MW}} = 0.008$~mag \citep{schlafly2011} but we do not correct for any host-galaxy extinction towards \name as multiple studies have shown it to be negligible \citep[e.g., ][]{patat2013, zhang2016}. M101 has several distance estimates in the literature due to its proximity \citep[e.g., ][]{matheson2012, jang2017, beaton2019}, but we adopt the distance of $6.4$~Mpc from \citet{shappee2011} to allow direct comparisons with previous studies. All phases are given relative to \tmax in the SN rest-frame.

\section{New and Archival HST Imaging}\label{sec:data}

\begin{deluxetable}{lccrc}
\tablecaption{New \hst WFC3 Observations \label{tab:hstinfo}}
\tablehead{
    \colhead{Filter} & \colhead{UT Date} & \colhead{Phase\tablenotemark{a}} & \colhead{Exp. Time} & \colhead{Vega Mag} \\ 
     & & [days] & [sec] & 
}
\startdata
\filtB & 2017-06-15 & 2103.5 & 6150  & $27.55\pm0.25$\\
\filtV & 2017-06-15 & 2103.5 & 5200 & $27.48\pm0.19$ \\
\filtR & 2017-06-17 & 2105.5 & 5400 & $27.18\pm0.32$\\
$F110W$ & 2017-06-17 & 2105.5 & 4010 & $26.73\pm0.35$\\
$F160W$ & 2017-06-17 & 2105.5 & 1910 & $24.70\pm0.18$\\
\filtB & 2018-03-30 & 2391.3 & 8950 & $27.89\pm0.45$\\
\filtV & 2018-03-26 & 2386.9 & 6300 & $28.07\pm0.24$\\
\filtR & 2018-03-26 & 2386.9 & 11500 & $27.50\pm0.37$\\
$F110W$ & 2018-03-30 & 2391.3 & 5610 & $27.02\pm0.37$\\
$F160W$ & 2018-03-30 & 2391.3 & 8010 & $25.19\pm0.22$\\ 
\enddata
\tablenotetext{a}{Relative to $t_{\rm{max}}$.}
\end{deluxetable}

\begin{figure*}
    \centering
    \includegraphics[width=0.9\linewidth]{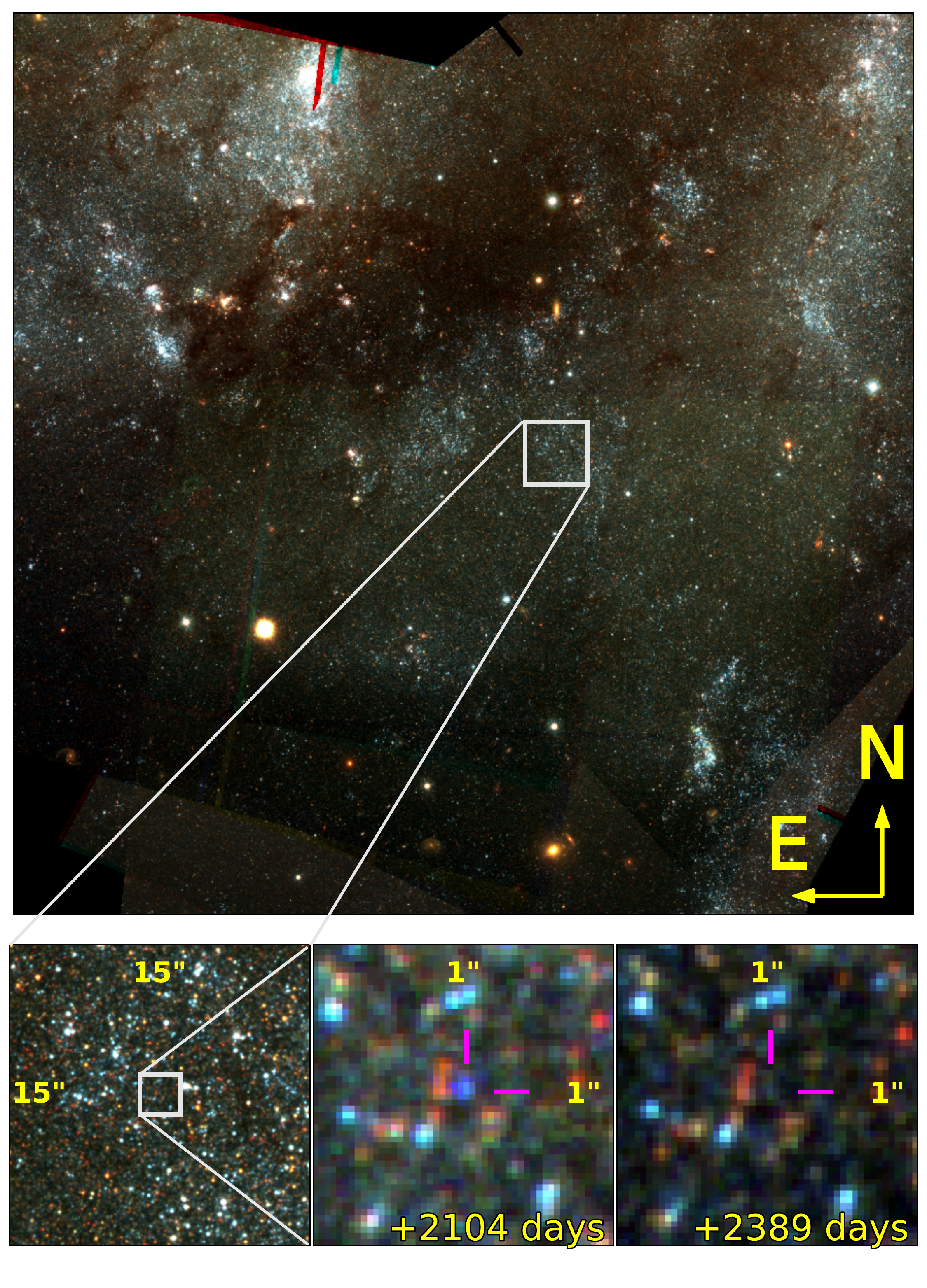}
    \caption{Composite HST color image of \name. The top and lower-left panels are comprised of all \hst observations. The lower-middle and lower-right panels are the single-epoch images centered on \name (marked with pink reticles) with the phase relative to \tmax given in the bottom right of each panel.}
    \label{fig:HSTrgb}
\end{figure*}

We obtained new photometry of \name at 2105 and 2390~days after maximum light using the UVIS and NIR modules of the Wide Field Planetary Camera 3 (WFC3) instrument on the Hubble Space Telescope (\hst). The new observations used the broadband \filtB, \filtV, \filtR, $F110W$, and $F160W$ filters (see Table \ref{tab:hstinfo}, GO-14678 and GO-15192, PI: Shappee). We also include 5 epochs of \hst photometry from \citet{shappee2017} which cover $\approx 1120-1840$~days after maximum light (GO-13737 and GO-14166, PI: Shappee) for the same optical and near-infrared (NIR) filters. 

We use the \textsc{dolphot} software package \citep{dolphin2000} to perform point spread function (PSF) fitting photometry on the individual images retrieved from the Mikulski Archive for Space Telescopes (MAST)\footnote{\url{mast.stsci.edu}}. We use the \texttt{*flc.fits} UVIS frames which are already corrected for charge transfer efficiency (CTE) losses and the \texttt{*flt.fits} NIR frames that do not need CTE corrections. Due to the faintness of \name in the new observations, we simultaneously analyze both old and new images with \textsc{dolphot} to ensure that the location of \name remains consistent across all epochs and filters. Individual frames are aligned to sub-pixel precision using the \textsc{TweakReg} software package with typical alignment errors of $\approx 0.1-0.2$ pixels. Then, \textsc{AstroDrizzle} is used to create deep (`drizzled') reference images for each filter with the drizzled \filtV image used as the \textsc{dolphot} reference frame. The composite color image is shown in Fig. \ref{fig:HSTrgb}. 

 Our $<2000$-day photometry is consistent with the published results of \citet{shappee2017} and \citet{kerzendorf2017}. The largest deviations between our photometry and the previously published results occur for the earliest $F438W$ observations, with a maximum discrepancy of $\approx 2\sigma$. We attribute this difference to updated charge transfer efficiency (CTE) corrections in the WFC3/UVIS pipeline, as the largest photometric discrepancies correspond to observations where 1) \name is far from the read-out amplifier and 2) the sky background is low enough ($< 10 e^-$ per image) for CTE corrections to dominate the uncertainty \citep[c.f., ][]{anderson2021}. While it is not possible to directly confirm that CTE is the source of the observed discrepancy, we note that the differences are at the $\lesssim 2\sigma$ level and the pseudo-bolometric light curve  derived in \S\ref{subsec:phot.bolLC} agrees with the pseudo-bolometric light curves calculated by \citet{shappee2017} and \citet{kerzendorf2017} to better than $1\sigma$.

\begin{figure*}
    \centering
    \includegraphics[width=\linewidth]{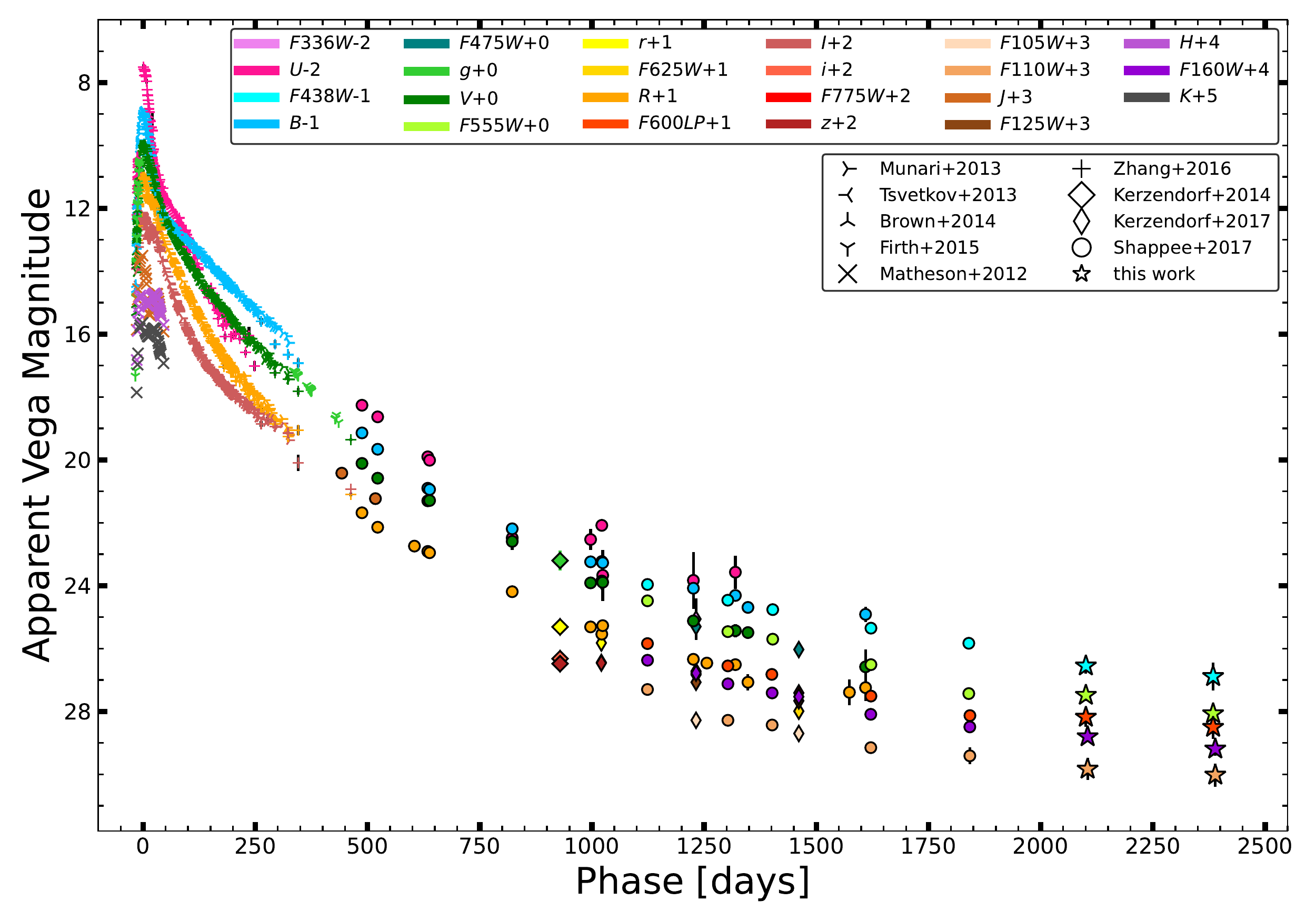}
    \caption{Complete optical and NIR light curve of \name.}
    \label{fig:fullLC}
\end{figure*}

\section{Pseudo-Bolometric Light Curve}\label{sec:phot}

Bolometric light curves provide important insight into the physical processes governing the observed emission. However, constructing a reliable bolometric light curve is challenging due to both observational factors and limits in our theoretical understanding of the dominant emission mechanisms. \S\ref{subsec:phot.bolLC} outlines our procedure for constructing the (pseudo-)bolometric light curve, including comparisons to previous works and assumptions about the mid-IR contribution. We fit simple radioactive decay models to the derived pseudo-bolometric light curve in \S\ref{subsec:phot.nuclsyn} and compare our isotopic ratios to other measurements in the literature. Finally, we discuss the effects of positron propagation within the ejecta and potential sources of contamination in \S\ref{subsec:phot.positrons} and \S\ref{subsec:phot.contam}, respectively. 

\subsection{Constructing the Pseudo-Bolometric Light Curve}\label{subsec:phot.bolLC}

\begin{figure}
    \centering
    \includegraphics[width=\linewidth]{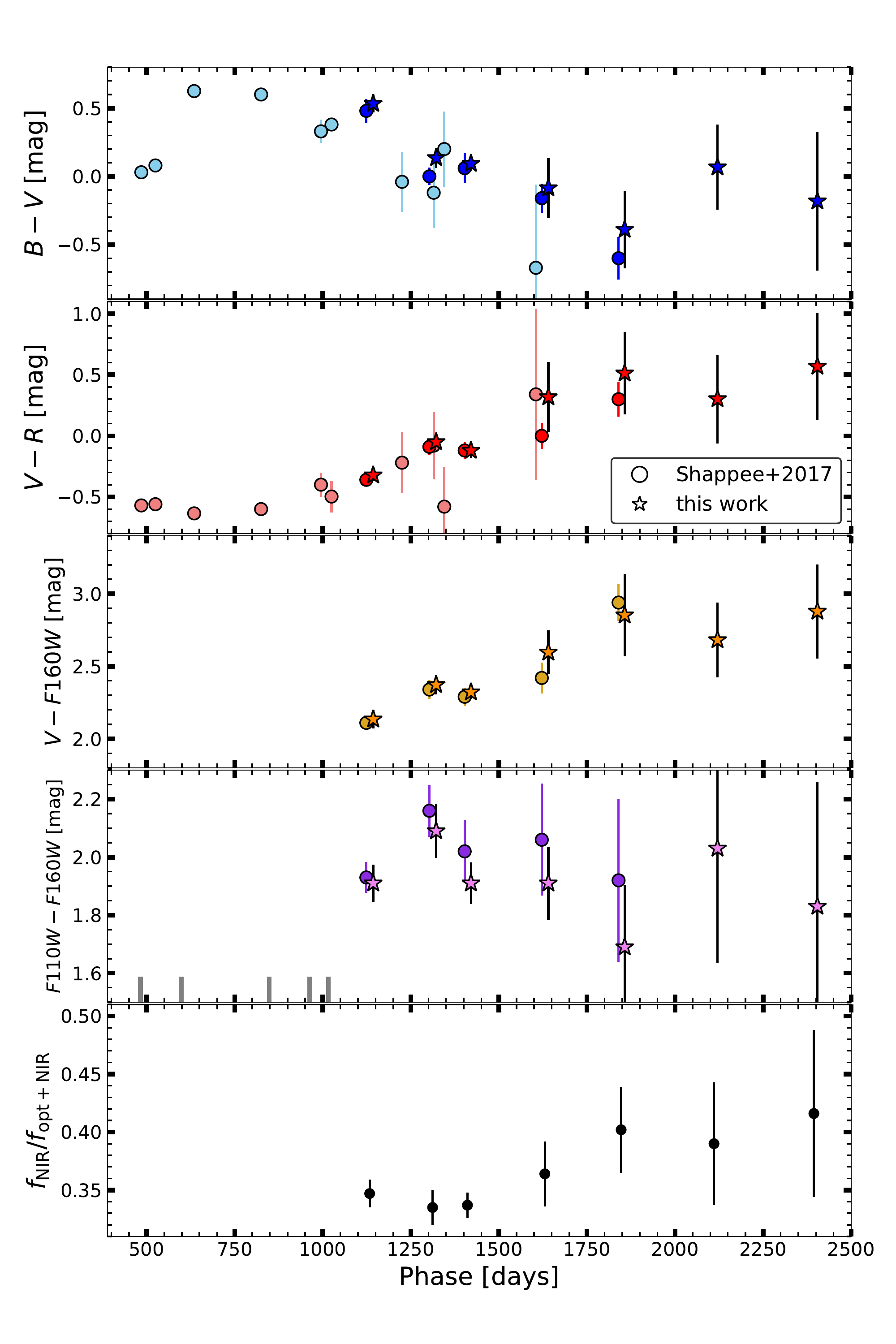}
    \caption{Color evolution of \name at $>450$~days including $B-V$, $V-R$, $V-F160W$, and $F110-F160W$. We assume the \hst \filtB, \filtV, and \filtR filters are roughly equivalent to the Johnson $BVR$ filters (e.g., Fig. 2 from \citealp{shappee2017}). Light-shaded markers represent Large Binocular Telescope (LBT) photometry whereas darker markers signify \hst photometry. The \hst photometry used in this work is offset in time by 20~days for visual clarity. Gray ticks along the lower axis mark spectral epochs from \citet{taubenberger2015}, \citet{graham2015}, and \citet{tucker2022}. The NIR fraction, defined as the ratio of NIR flux to optical + NIR flux, is shown in the bottom panel.}
    \label{fig:color-color}
\end{figure}

\begin{figure}
    \centering
    \includegraphics[width=\linewidth]{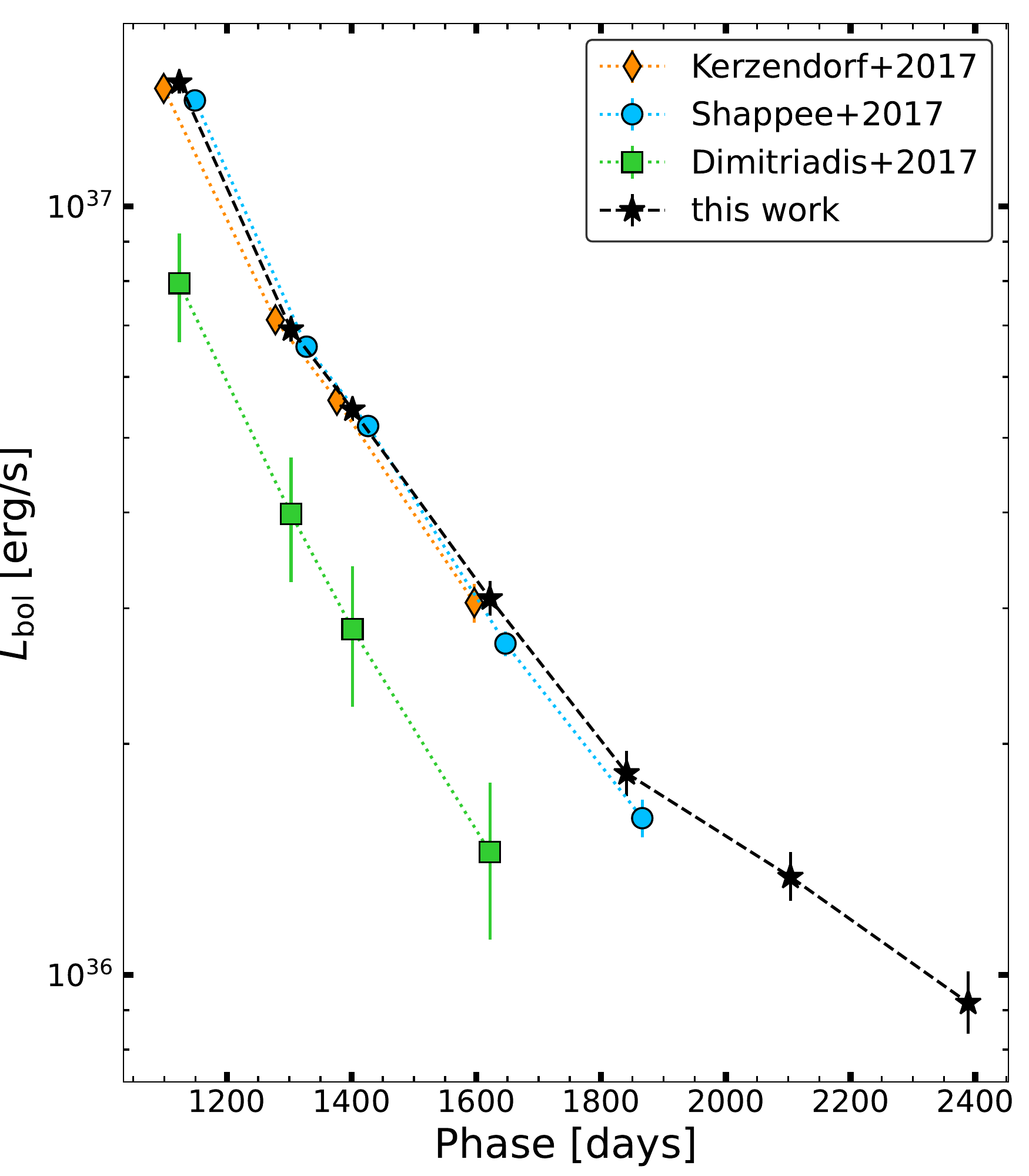}
    \caption{Comparison between our pseudo-bolometric light curve (black) and \citet{kerzendorf2017}, \citet{shappee2017}, and \citet{dimitriadis2017}.}
    \label{fig:bolcompare}
\end{figure}

Our observations cover optical and NIR wavelengths ($0.4-1.7~\mu\rm m$). Both the $+1000$-day theoretical spectrum from \citet{fransson2015} and the optical + NIR photometry from \citet{shappee2017} predict a NIR fraction\footnote{We use the same formalism for the fractional NIR contribution presented by \citet{dimitriadis2017}: $F_{\rm{NIR}} = L_{\rm{NIR}} / (L_{\rm{NIR}} + L_{\rm{opt}})$.} of $F_{\rm{NIR}}\approx 35\%$ which agrees with our results in the bottom panel of Fig. \ref{fig:color-color}. Our observations are consistent with a flat or modestly increasing NIR fraction at $\gtrsim 1500$~days after \tmax with our latest observations reaching $F_{\rm{NIR}} = 41\pm 7\%$.

Because our observations are limited to optical and NIR wavelengths, we must make assumptions about the mid-IR (MIR) contribution. \citet{axelrod1980} predicted an ``infrared catastrophe'' (IRC) beginning $\sim 500$~days after explosion due to the diminishing temperature and density in the ejecta being unable to populate the optical \fFe{2} and \fFe{3} transitions. In the IRC picture, the late-time emission should be dominated by fine-structure line emission in the MIR with no optical emission. Empirically, however, there is no evidence for the IRC \citep[e.g., ][]{stritzinger2007, hristov2021} and the most recent theoretical modeling from \citet{fransson2015} show that the IRC does not set in by day 1000 due to the redistribution of UV flux through multiple scatterings and fluorescence. 

The fraction of the bolometric luminosity emitted in the mid-IR fine-structure lines remains unclear. Recent observational (e.g., \citealp{black2016}, \citealp{mazzali2020}, \citealp{tucker2022}) and theoretical \citep[e.g., ][]{wilk2020} studies have found tentative evidence for clumping in the ejecta of \sneia which has important ramifications for the presence of the IRC. The degree of clumping determines where energy is deposited and affects both the observed spectra and the MIR emission. Fine-structure cooling will occur only in the low-density regions in a clumpy medium. Due to these uncertainties, we include the fraction of total luminosity emitted in the optical and NIR as a parameter constrained by a prior in our subsequent analyses.

We compute the pseudo-bolometric light curve by warping \citep[e.g., ][]{hsiao2007} the theoretical +1000-day spectrum from \citet{fransson2015} to match the observed photometry and then integrating the optical and NIR flux. The pseudo-bolometric light curve is unaffected by our choice to use the theoretical spectrum compared to the observed +1000-day spectrum from \citet{taubenberger2015} with a NIR correction. The uncertainty in the distance to \name/M101 is not propagated to the uncertainty in the pseudo-bolometric light curve as we do not attempt to derive true isotopic masses but only constrain the relative isotopic abundances in the following analyses. However, we do include a 5\% systematic uncertainty when estimating our final pseudo-bolometric light curve to account for our incomplete knowledge regarding the evolution of the MIR emission. This systematic uncertainty is a sub-dominant source of error at $\gtrsim1600$~days after \tmax compared to the statistical photometric uncertainties. 

The larger pixel scale of the WFC3/IR module and the very red stars near \name make crowding a potential source of contamination at these epochs (i.e., Fig. 1 from \citealp{shappee2017}). To ensure contaminated NIR photometry does not dictate our results, we re-derive the pseudo-bolometric light curve by warping and integrating the optical portion of the spectrum then including a constant NIR fraction of $35\pm5\%$ (e.g., Fig. 7 from \citealp{dimitriadis2017}). The resulting pseudo-bolometric light curve is consistent at $<1\sigma$ from the procedure including the NIR photometry (see the bottom panel of Fig. \ref{fig:color-color}), indicating that the NIR photometry is likely reliable.

Our final pseudo-bolometric light curve is compared to the results from \citet{shappee2017}, \citet{kerzendorf2017}, and \citet{dimitriadis2017} in Fig. \ref{fig:bolcompare}. Our pseudo-bolometric light curve is slightly higher than previous estimates from \citet{shappee2017} and \citet{kerzendorf2017} due to slightly higher photometric measurements (see \S\ref{sec:data}) but consistent within the uncertainties. We note that the pseudo-bolometric light curve derived by \citet{dimitriadis2017} disagrees with our results and those of \citet{shappee2017} and \citet{kerzendorf2017}. This discrepancy is likely due to the different methodology employed by \citet{dimitriadis2017} who construct a pseudo-bolometric light curve from $R$-band photometry and a NIR correction instead of integrating over broadband multi-filter photometry as done in this work, \citet{shappee2017}, and \citet{kerzendorf2017}. The main spectral feature in the $R$-band at $\gtrsim 500$~days after \tmax is \CaII \citep{tucker2022}. The $4000-6000$~\AAA wavelength range is dominated by a complex blend of forbidden iron lines powered by UV scattering which generally tracks the bolometric evolution \citep[e.g., ][]{fransson2015}. However, the ratio of \CaII flux to $4000-6000$~\AAA flux is not constant with time \citep{tucker2022}. This indicates that the \CaII feature (and thus $R$-band photometry in general) is not a good proxy for the bolometric evolution.

\vspace{1cm}
\subsection{Interpreting the Pseudo-Bolometric Light Curve}\label{subsec:phot.nuclsyn}

\begin{figure}
    \centering
    \includegraphics[width=\linewidth]{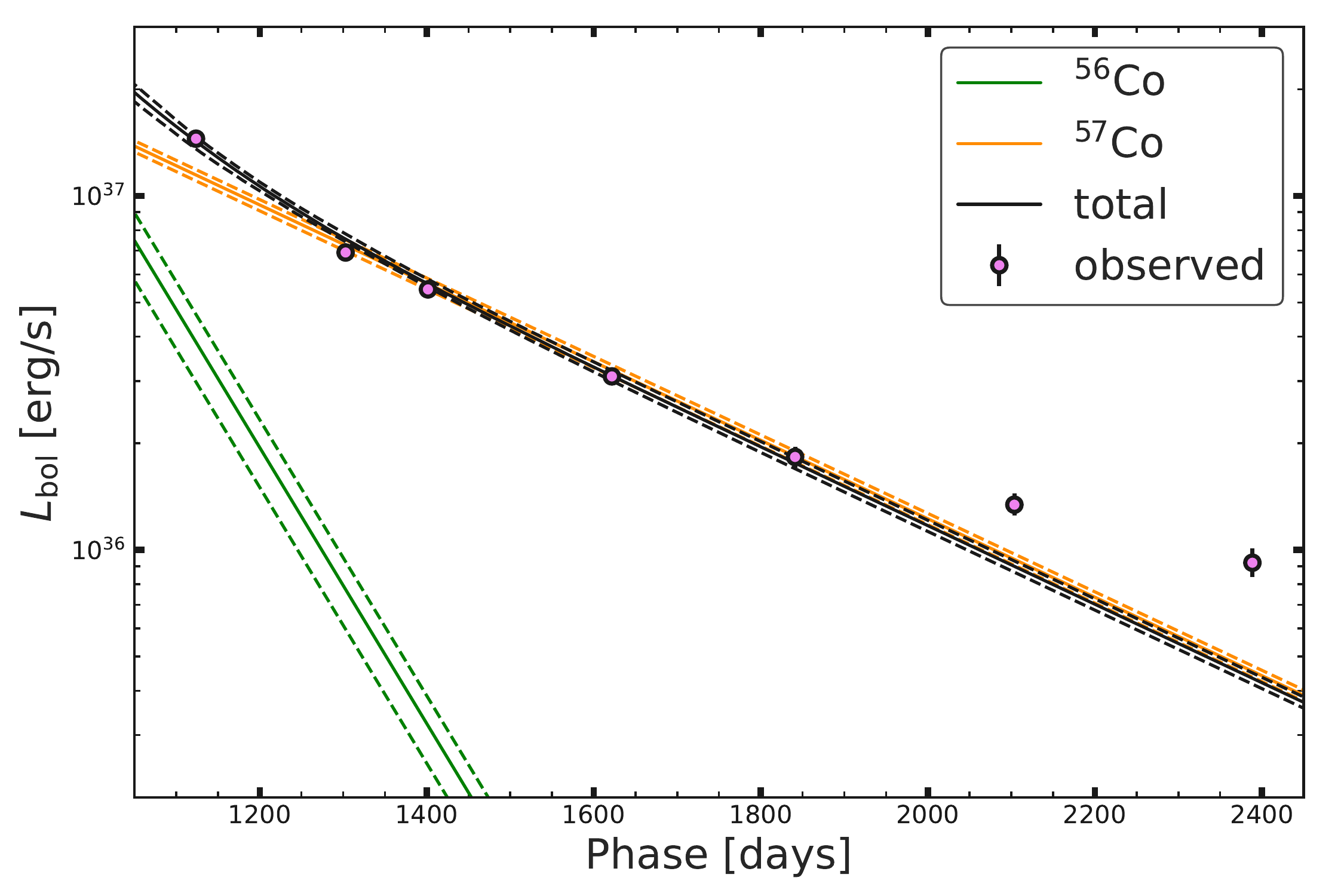}
    \includegraphics[width=\linewidth]{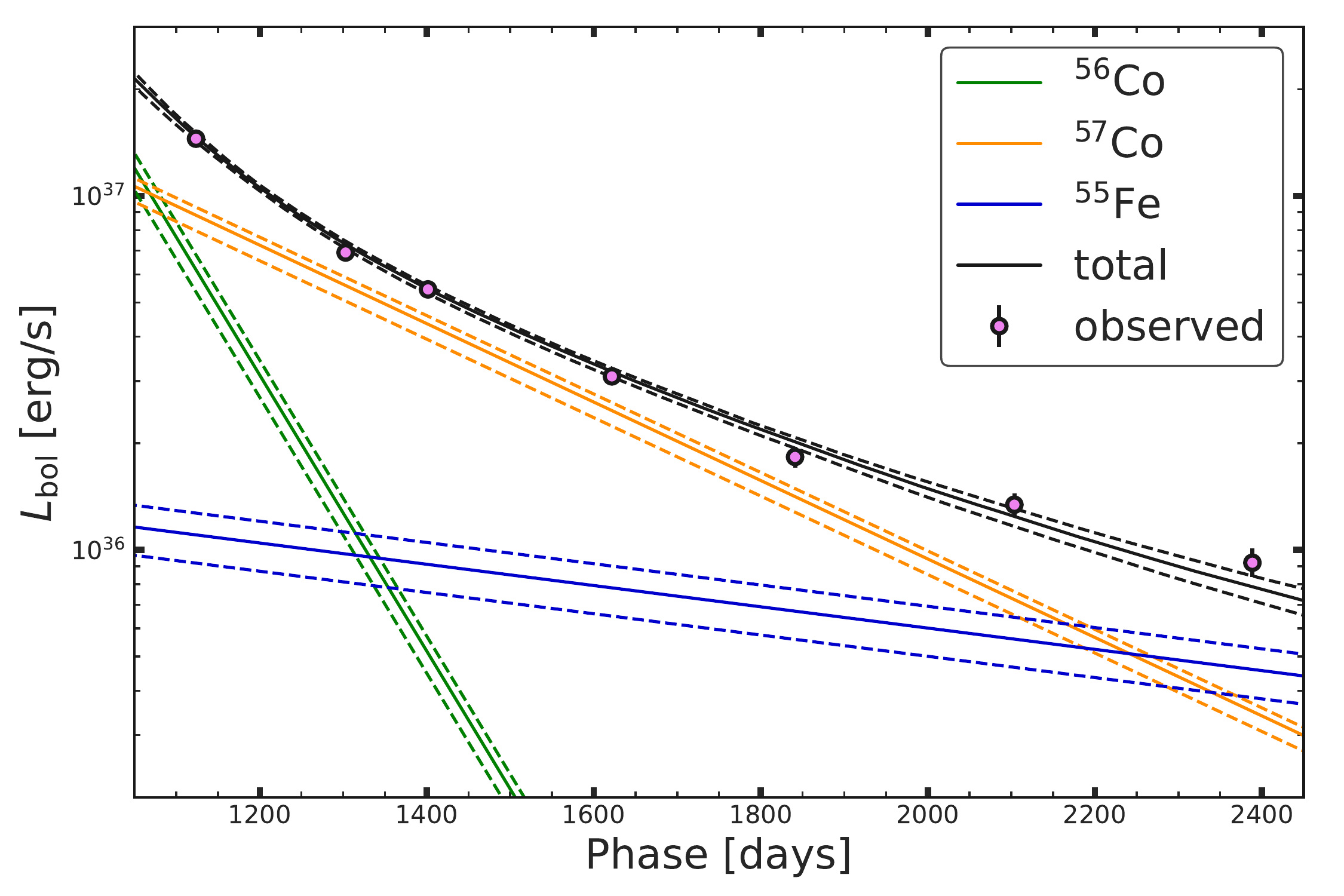}
    \includegraphics[width=\linewidth]{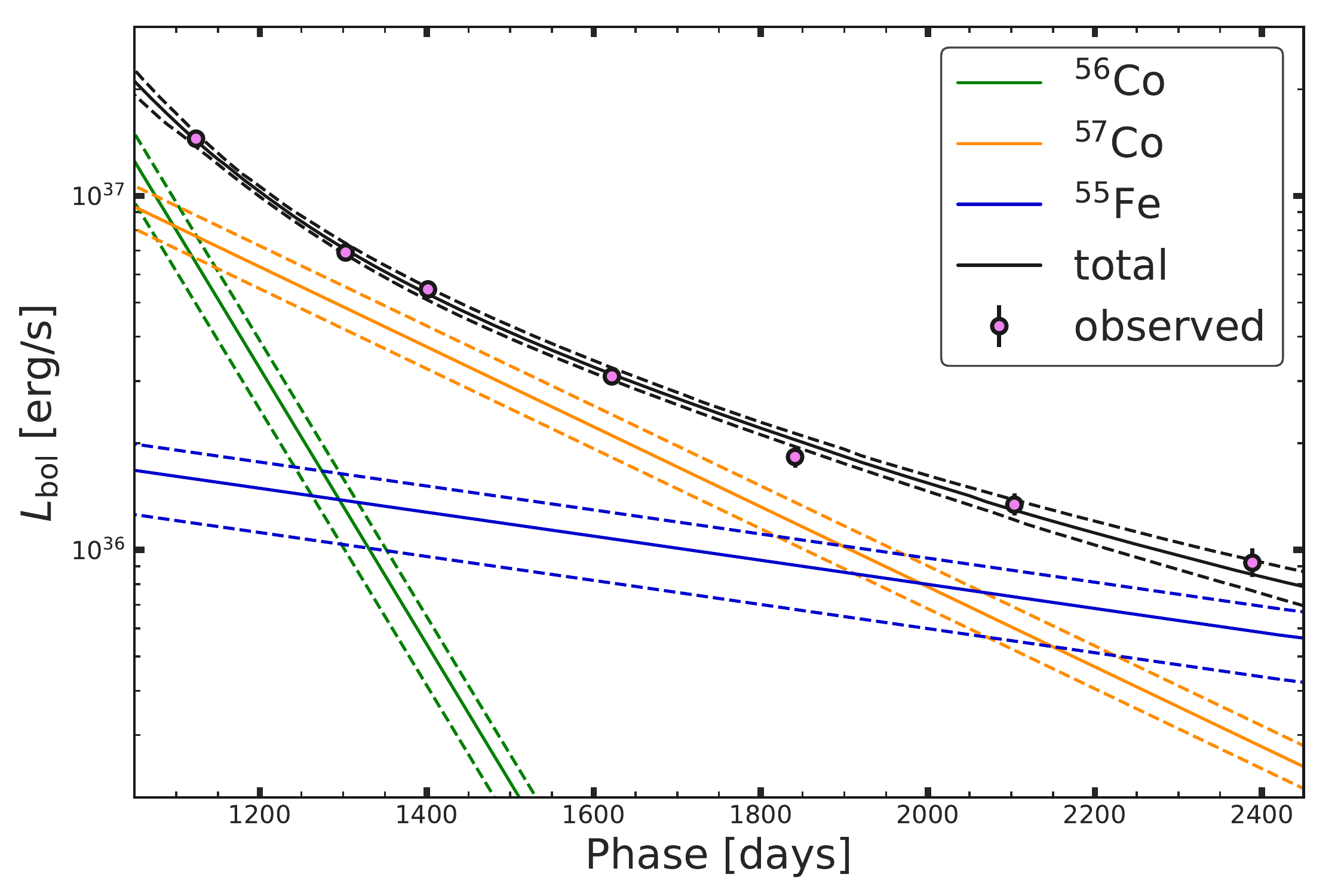}
    \caption{Fits to the $>1000$~day pseudo-bolometric light curve including energy input from \iCo{56} + \iCo{57} (top) and \iCo{56}+\iCo{57}+\iFe{55} (middle). The bottom panel shows the fit including all three radioactive decay chains and incorporating time-dependent X-ray escape. Solid and dashed lines represent the median and 68\% confidence intervals. The evolution of the pseudo-bolometric light curve can be completely explained with just energy input from radioactive decays.}
    \label{fig:decayfit}
\end{figure}

We model the bolometric light curve with energy input from the radioactive decays in Eq. \ref{eq:decaychains}. The decay chain of $^{44}\rm{Ti}\rightarrow^{44}\rm{Sc}\rightarrow^{44}\rm{Ca}$ is omitted for most of the analysis due to the long half-life ($\approx 60~\rm{years}$) of the first decay and the very small mass of \Ti expected from \sneia \citep[ $\lesssim 10^{-5}~M_\odot$; e.g., ][]{iwamoto1999, seitenzahl2009}. We do provide simple constraints on the \Ti mass in \S\ref{subsec:phot.nuclsyn}. For simplicity, we refer to the masses of radioactive isotopes and their daughter nuclei by their mass number (i.e., $M_{56} = M_{56\rm{Ni}} + M_{56\rm{Co}} + M_{56\rm{Fe}}$).\footnote{These values only represent radioactive species and do not account for directly-synthesized stable isotopes. The only stable isotope produced in enough quantities for this to matter is stable \iFe{56}.} At $> 1000$~days after \tmax, \gamrays are free-streaming and do not contribute meaningfully to the emitted energy. However, the efficiency of X-ray trapping is less clear, especially at the epochs where \iFe{55} decays dominate the total energy deposition. Two simple models are considered in the subsequent analysis. The first assumes all X-rays are trapped and deposit their energy locally, and the second uses a simple prescription for time-dependent X-ray escape based on the models of \citet{seitenzahl2015}, while still assuming the trapped X-rays deposit their energy locally in the ejecta.

The second decay of the first decay chain, ${^{56}\rm{Co}\rightarrow^{56}\rm{Fe}}$, occurs primarily by electron-capture, but 19\% of these are through $\beta^+$ decay \citep{TabRad_v0, TabRad_v8}, producing a high-energy ($\sim 1$~MeV) positron ($e^+$). For many years it was thought that a small fraction of these positrons would escape into the interstellar medium \citep[$\sim 1-10\%$; e.g., ][]{milne1999, milne2001}. However, the inclusion of NIR photometry suggests complete positron trapping \citep[e.g., ][]{sollerman2004, stritzinger2007} and more recent studies have yet to find any evidence for \eplus escape \citep[e.g., ][]{hristov2021}. For these reasons we assume the \iCo{56} positrons are completely trapped in the ejecta, although we discuss the time-dependent energy deposition from \eplus in \S\ref{subsec:phot.positrons}.

There are theoretical predictions \citep{axelrod1980, fransson2015} for emission at $>2.5~\mu\rm m$ (see \S\ref{subsec:phot.bolLC}), so our initial model includes energy input from \iCo{56} + \iCo{57} and assumes some fixed fraction $f_{\rm{opt}+\rm{NIR}}$ of the true bolometric luminosity is emitted within $0.3-2.5$~\um. The effect of \iCo{57}/\iNi{57} on the late-time light curves of \sneia has been invoked in the literature for both \name \citep[e.g., ][]{fransson2015, shappee2017, kerzendorf2017, dimitriadis2017} and other \sneia \citep[e.g., ][]{graur2016, flors2018, jacobson-galan2018, yang2018}. Contribution from the decay of \iFe{55} is excluded (i.e., $M_{55} \equiv 0~M_\odot$) for this initial model and we set simple priors on the parameters, including ${M_{57} > 0~M_\odot}$, $M_{56} = 0.5\pm0.1~M_\odot$, and ${0 \leq f_{\rm{opt}+\rm{NIR}} \leq 1}$. The distance to M101 of $6.4\pm0.4$~Mpc from \citet{shappee2011} is included as a prior due to the covariance between the distance and \M{56}. The prior on \M{56} is derived from the \M{56} values reported by \citet{pereira2013} and \citet{mazzali2015} using separate methods and data\footnote{We do not include the \iNi{56} mass derived by \citet{scalzo2014}, as they use the same SuperNova Factory observations as \citet{pereira2013}. The \iNi{56} mass derived by \citet{childress2015} is excluded because they explicitly use \name as an anchor for deriving their relation. }. We note that removing the prior on $M_{56}$ does not affect the final conclusions but increases the resulting uncertainties (i.e., broadens the parameter posterior distributions). For the model excluding contributions from \iFe{55}, we find $\log_{10}(M_{57}/M_{56}) = -1.36^{+0.12}_{-0.09}$ and $f_{\rm{opt}+\rm{NIR}} = 0.34^{+0.07}_{-0.09}$, shown in the left panel of Fig. \ref{fig:decayfit}. The observations at $>2000$~days after explosion are poorly fit by this model.

In practice, \iFe{55} is expected to dominate the energy input into the ejecta at $\gtrsim 1800$~days after \tmax \citep[e.g., ][]{seitenzahl2009, kushnir2020} but it has never previously been detected in the light curve of a \snia. If we now add \iFe{55} with $M_{55} > 0~M_\odot$, the same priors on the other three variables, and assume all X-rays are confined to the ejecta, we find $\log_{10} (M_{57}/M_{56}) = -1.69^{+0.10}_{-0.09}$, $\log_{10}(M_{55}/M_{57}) = -0.66^{+0.10}_{-0.12}$, and $f_{\rm{opt}+\rm{NIR}} = 0.68^{+0.17}_{-0.16}$. This model provides an excellent fit to the observations. For a fiducial \M{56} mass of $0.5~\rm M_\odot$, these ratios correspond to isotopic masses of \M{57}$\approx 0.01~\rm M_\odot$ and \M{55}$\approx 0.002~\rm M_\odot$.

If X-rays are allowed to escape the ejecta, then some of the energy released by the decay of \iFe{55}$\rightarrow$\iMn{55} will not be deposited into the ejecta and thus not contribute to the bolometric light curve. To incorporate X-ray escape, we scale the emitted luminosity for each decay chain by $1- X_i f(t)$ where $X_i$ represents the fraction of decay energy released as X-ray photons ($\approx1\%$ for \iCo{56}, $\approx 17\%$ for \iCo{57}, and $\approx 29\%$ for \iFe{55}, see Table 1 from \citealp{seitenzahl2009}) and $f(t)$ is the time-dependent X-ray trapping function. \citet{seitenzahl2015} estimate $10\%$ of the X-rays escape at 1000~days after explosion which increases linearly to $\sim50\%$ at day 2400. Even assuming the explosion model with a higher X-ray escape fraction, the escaping X-rays constitute $<15\%$ of the total decay energy at 2400~days and fitting the pseudo-bolometric light curve with this model produces similar results to the model assuming complete X-ray trapping, as shown in the bottom panel of Fig. \ref{fig:decayfit}. Including time-dependent X-ray escape produces $\log_{10}(M_{57}/M_{56}) = -1.76^{+0.13}_{-0.12}$, $\log_{10}(M_{55}/M_{57}) = -0.45^{+0.13}_{-0.14}$ and $f_{\rm{opt+NIR}} = 0.69^{+0.17}_{-0.17}$. For a fiducial \M{56} mass of $0.5~M_\odot$, these isotope ratios correspond to synthesized masses of $M_{57} \approx 0.01~M_\odot$ and $M_{55} \approx 3\times10^{-3}~M_\odot$.

\sneia are not thought to produce significant amounts of \Ti \citep[e.g., ][]{iwamoto1999, hoeflich2017} but the resurgence of explosion models invoking He burning shells \citep[e.g., ][]{shen2021} may challenge this assumption. Therefore, we check if the inclusion of \decaychainfour affects our results. The first decay, \Ti$\rightarrow$\Sc, has a $\approx 60$~year half-life but $\approx 94\%$ of the subsequent ${^{44}\rm{Sc}\rightarrow^{44}\rm{Ca}}$ decays produce a high-energy ($\approx 1.4$~MeV) positron within a few hours \citep[e.g., ][]{seitenzahl2009, nucdata44}. A synthesized \Ti mass of $10^{-4}~M_\odot$ supplies enough decay energy ($\approx 9\times 10^{35}~\rm{erg}~\rm s^{-1}$, \citealp{seitenzahl2014}) to match our last epoch of \hst photometry (cf. Fig. \ref{fig:decayfit}). We do not fit for the amount of synthesized \Ti in our pseudo-bolometric light curve for two reasons. The first is practical, as the near-constant luminosity from \M{44} is degenerate with \M{55} similar to including luminosity from a surviving companion (cf. \S\ref{subsubsec:phot.donor}). The second is theoretical, as we cannot assume that \fopt is the same for the inner and outer ejecta and \Ti would be primarily produced during surface He burning. With these caveats, we place a rough upper limit of $M_{44} < 10^{-3}~M_\odot$ on the synthesized \Ti mass which would supply the entire pseudo-bolometric luminosity at 2400~days assuming $f_{\rm{opt}+\rm{NIR}} = 0.1$.

Finally, we compare our \M{57}/\M{56} ratio to previous estimates. \citet{shappee2017} analyzed the first $\sim 1100-1800$~days of \name and report a \M{57}/\M{56} mass ratio of $\log_{10}(M_{57}/M_{56}) = -1.59^{+0.06}_{-0.07}$, in good agreement with ours. They did not have a clean detection of \iFe{55} but placed a 99\% upper limit at $\log_{10}(M_{55}/M_{57}) < -0.66$, broadly consistent with our new estimate. Our \M{57}/\M{56} ratio also agrees reasonably well with estimates by \citet{graur2016},  \citet{jacobson-galan2018}, and \citet{flors2018} for other \sneia. However, our results disagree with the results of \citet{dimitriadis2017} for \name and with \citet{yang2018} for SN~2014J. The results of \citet{dimitriadis2017} are likely based on an inaccurate pseudo-bolometric light curve, as discussed in \S\ref{subsec:phot.bolLC}. The pseudo-bolometric light curve derived by \citet{yang2018} for SN~2014J only extends to $\approx 1200$~days after \tmax where \iCo{56} still dominates the light curve (c.f. Fig. \ref{fig:decayfit}). Additionally, \citet{yang2018} assume a fixed \M{55}/\M{57} ratio of $\log_{10}(M_{55}/M_{57}) \approx -0.1$. This is consistent with some of the explosion models shown in Fig. \ref{fig:yields} but the resulting \M{57}/\M{56} ratio of $\log_{10}(M_{57}/M_{56}) = -1.18^{+0.03}_{-0.04}$ is highly inconsistent with both our result and the bulk of \snia explosion models (e.g., Fig. \ref{fig:yields}). 

Our results do not require supersolar abundances of iron-group elements (IGEs) as suggested by some studies in the literature \citep[e.g., ][]{graur2016}. In fact, our isotopic constraints agree at $\lesssim 2\sigma$ with the Solar abundances from \citet{asplund2009} with the caveat that the \iFe{56} abundance inferred from the Sun includes both stable \iFe{56} and daughter \iFe{56} from radioactive \iNi{56}. While a single object does not lend itself well to statistical inferences, we can at least say that \name, the quintessential \snia, produced $M_{55,56,57}$ abundance ratios in rough agreement with the Solar values.

\subsection{Positron Propagation}\label{subsec:phot.positrons}

\begin{figure}
    \centering
    \includegraphics[width=\linewidth]{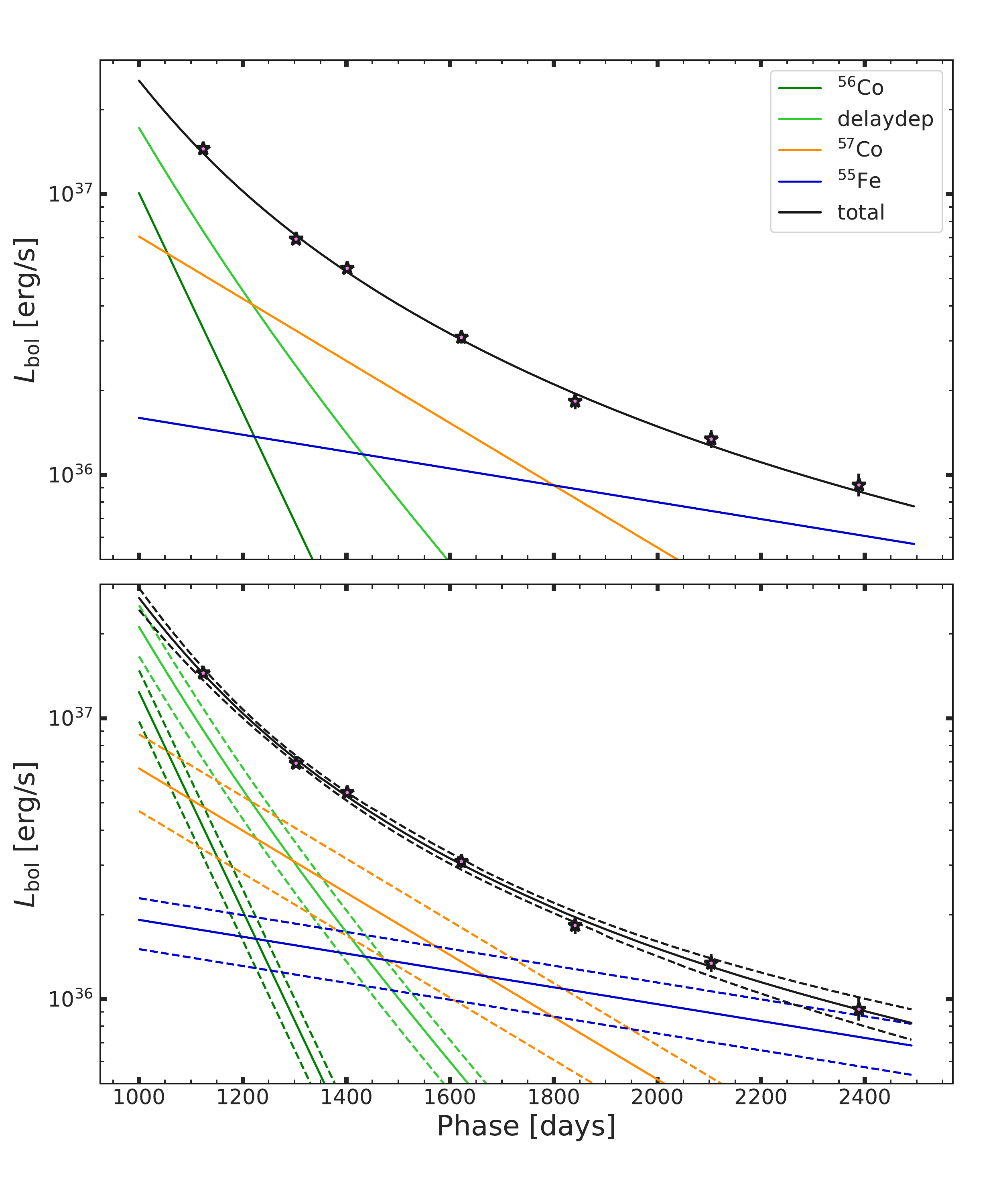}
    \caption{Fits to the pseudo-bolometric light curve when including the ``delayed deposition'' effect \citep{kushnir2020} assuming Solar values for the isotopic ratios (top panel) and allowing the isotopic ratios to vary (bottom panel). The color scheme is the same as in Fig. \ref{fig:decayfit} with the addition of the delayed deposition effect (light green).}
    \label{fig:delaydep}
\end{figure}

Although we assume the \iCo{56} positrons cannot escape the ejecta (\S\ref{subsec:phot.bolLC}), there are other possible effects on the light curve from the propagation of positrons in the ejecta. As the ejecta expands and the density decreases, it will take longer for a positron to deposit its energy into the surrounding ejecta (i.e., ``delayed deposition''; \citealp{kushnir2020}). This effect extends the energy deposition from \iCo{56} and slows the decline rate compared to pure radioactive decay. It is unclear at what level, if any, this process will affect the observed light curve due to uncertainties regarding magnetic field generation within the ejecta (e.g., \citealp{hristov2021,gupta2021}) and the degree of clumping in the iron core \citep[e.g., ][]{black2016, wilk2020}. With these caveats, we briefly consider the potential effect of delayed deposition on the derived isotope ratios.

We first test the model of \citet{kushnir2020} where the isotopic ratios are set to the Solar values, and \fopt and the mass-weighted density, \rhot, are free parameters. We set a simple prior on \rhot of $-1 < \log_{10}( \langle \rho \rangle t^3) < 1$ (see Fig. 2 from \citealp{kushnir2020}) and the same prior on \fopt as in \S\ref{subsec:phot.nuclsyn}. This model produces similar results to \citet{kushnir2020}, namely $\langle\rho\rangle t^3 = 0.14^{+0.02}_{-0.01}$ and $f_{\rm{opt}+\rm{NIR}} = 0.39 ^{+0.02}_{-0.02}$ which is shown in the top panel of Fig. \ref{fig:delaydep}.

Next, we test a model where the isotopic ratios and \fopt are allowed to vary but \rhot is set $0.14$, the value derived for \name by \citet{kushnir2020}. The first thing to note is that our detection of \iFe{55} is robust to the inclusion of the delayed deposition effect, as most of the \iCo{56} \eplus have completely deposited their energy into the ejecta by the time \iFe{55} dominates the light curve (c.f. Fig. 3 from \citealp{kushnir2020}). However, the \iNi{57} value does change as energy originally attributed to \iNi{57} is instead attributed to surviving \iCo{56}. This results in a similar \M{57}/\M{56} ratio of $\log_{10}(M_{57}/M_{56}) = -1.77^{+0.20}_{-0.22}$~dex but increases \fopt and the \M{55}/\M{57} ratio to $f_{\rm{opt}+\rm{NIR}} = 0.46^{+0.16}_{-0.12}$ and $\log_{10}(M_{55}/M_{57}) = -0.20^{+0.22}_{-0.20}$, respectively. The delayed energy deposition from \iCo{56} \eplus essentially mimics the evolution of \iNi{57}, as shown in the bottom panel of Fig. \ref{fig:delaydep}. 

While the model including delayed deposition produces isotope ratios in agreement with our previous estimates, this effect is in need of further study before adopting it in our analysis. The mass-weighted density \rhot predicted from explosion models is in slight tension with observations (e.g., Fig. 2 from \citealp{kushnir2020}). Additionally, it is unclear how clumping of the ejecta affects \rhot which dictates the strength of the delayed deposition effect.

\subsection{Potential Sources of Contamination}\label{subsec:phot.contam}

\begin{figure}
    \centering
    \includegraphics[width=\linewidth]{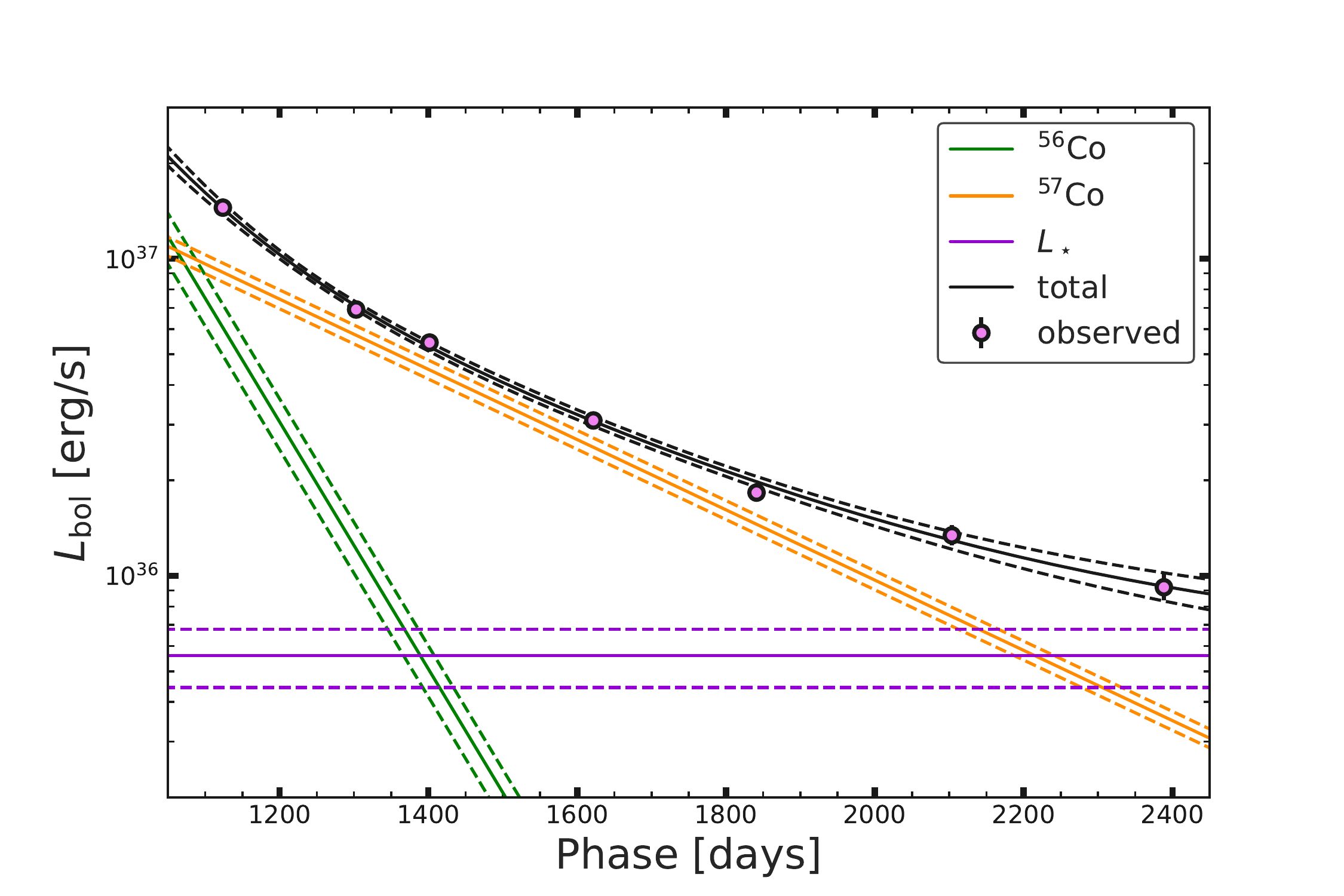}
    \caption{Model fit including luminosity from \iCo{56}, \iCo{57}, and a surviving companion star ($L_\star$). \iFe{55} is excluded from this model (see \S\ref{subsec:phot.contam}).}
    \label{fig:lstar}
\end{figure}

Finally, we consider plausible sources of contamination to the pseudo-bolometric light curve. None of the HST images show any evidence for a resolved light echo. The late-time color evolution strongly disfavors the presence of an unresolved light echo \citep{shappee2017, dimitriadis2017}, consistent with \name exploding in an extremely clean ISM environment \citep[e.g., ][]{patat2013, zhang2016}. However, we do consider contributions from a surviving post-impact donor star in \S\ref{subsubsec:phot.donor} and a breakdown in the ionization and recombination timescales in \S\ref{subsubsec:phot.freezeout}.

\subsubsection{A Surviving Post-Impact Companion Star}\label{subsubsec:phot.donor}

Another potential luminosity source at these epochs is a surviving donor star impacted by the explosion \citep[e.g., ][]{marietta2000, pan2012a, boehner2017}. The kick velocity imparted to the companion star is too low ($\lesssim 200$~\kms, \citealp{pan2012a, liu2013, boehner2017}) to produce a resolved shift in the \hst images. However, the injection of energy into the stellar envelope should produce a significant increase in the photospheric temperature and luminosity \citep[e.g., ][]{podsiadlowski2003, pan2012b, shappee2013b}. When the luminosity increases is unclear, as some models predict a rapid ($\lesssim 1$~year) thermalization of the stellar envelope and a subsequent rapid increase in luminosity \citep[e.g., ][]{shappee2013b, pan2013} yet other models predict the thermalization takes $\gtrsim 100$~years \citep[e.g., ][]{pan2012b, liu2021}. Thus, we test a few simple models where some of the pseudo-bolometric luminosity is provided by luminosity from a surviving donor star (\Lstar). 

Fig. \ref{fig:lstar} shows that our detection of \iFe{55} can be replaced by the constant luminosity from a surviving companion with $L_\star = 150\pm30~L_\odot$ and similar values for $\log_{10}(M_{57}/M_{56})$ and \fopt. Including luminosity input from both a surviving companion and \iFe{55} as free parameters results in sub-threshold ($<3\sigma$) detections for both components, with best-fit parameters $L_\star = 130^{+40}_{-50}~L_\odot$ and $\log_{10}(M_{55}/M_{57}) = -1.8\pm0.8$~dex. The other parameters (\fopt and $M_{57}/M_{56}$) are unaffected by the inclusion of a surviving companion. 

The detection of \iNi{57} without the detection of \iFe{55} would challenge most \snia explosion models (c.f. Fig. \ref{fig:yields}), so we consider a surviving companion star dominating the late-time light curve unlikely. However, this does not exclude models where it takes $\gtrsim 6.5$~years for the stellar envelope to thermalize (e.g., the main-sequence star models of \citealp{pan2012b} and the He-star models of \citealp{liu2021}). Interestingly, there are also models for both main-sequence stars \citep{shappee2013b} and He-stars \citep{pan2013} predicting a companion luminosity far above our detection. For this reason we conclude any surviving post-impact companion star must have $L_\star \ll 100~L_\odot$ at $\leq 6.5$~years after explosion to prevent introducing discrepancies with \snia explosion models but the exact type of star(s) excluded by this result is model-dependent. 

\subsubsection{Ionization Freeze-Out}\label{subsubsec:phot.freezeout}

Finally, we briefly discuss ionization ``freeze-out'' where the balance between ionization and recombination breaks down \citep[e.g., ][]{fransson1993}. This occurs at low densities and results in a flattening of the late-time light curve as ions become ``frozen'' due a lack of free electrons for recombination, essentially storing energy and delaying the timescale between energy injection and emission. \citet{kerzendorf2017} argue this is one of the reasons no conclusions can be drawn about isotopic yields based on the late-time light curve of \name. They show a fit to the $\approx 900-1600$~day pseudo-bolometric light curve using a simple prescription for freeze-out and show it can allow $M_{\rm{57}} \equiv 0~M_\odot$.

Approximating the impact of freeze-out as an additional luminosity source that evolves as $L \propto t^{-3}$ (as done by \citealp{kerzendorf2017}) can still produce a good fit to the late-time light curve. However, this speaks more to the flexibility of the model than its reliability. Luminosity from either \iNi{57} or \iFe{55} can be replaced with the $t^{-3}$ model for freeze-out, as the radioactive decay slopes are close enough to $-3$ and the additional parameter for scaling the freeze-out contribution provides plenty of flexibility in the model.

With these considerations in mind, we briefly discuss the applicability of freeze-out to \name at these epochs (see \citealp{fransson1993} for a discussion). Freeze-out sets in for the highest-velocity and lowest-density regions first. \citet{fransson2015} show the +1000-day optical spectrum is mainly comprised of \ion{Fe}{1} emission. Non-Fe emission components (mainly Ca) contribute a diminishing fraction of the luminosity with time \citep{tucker2022}. The Fe-dominated emission means that the luminosity is dominated by the iron-rich core instead of the lower-density, higher-velocity outer ejecta. \citet{fransson2015} state that the density and temperature of the inner core of \name are very similar to the inner core of SN~1987A modeled by \citet{jerkstrand2011}. Furthermore, \citet{jerkstrand2011} state that time-dependent effects from freeze-out have not set in for the iron-rich inner core of SN~1987A even $\sim 8$~years after explosion. While there are obvious differences in composition and density profiles between the cores of SN~1987A and \name, it suggests that freeze-out is likely a subdominant contributor to the late-time emission and that the derived isotopic ratios are valid. 

\section{Discussion}\label{sec:discuss}

\begin{figure*}
    \centering
    \includegraphics[width=\linewidth]{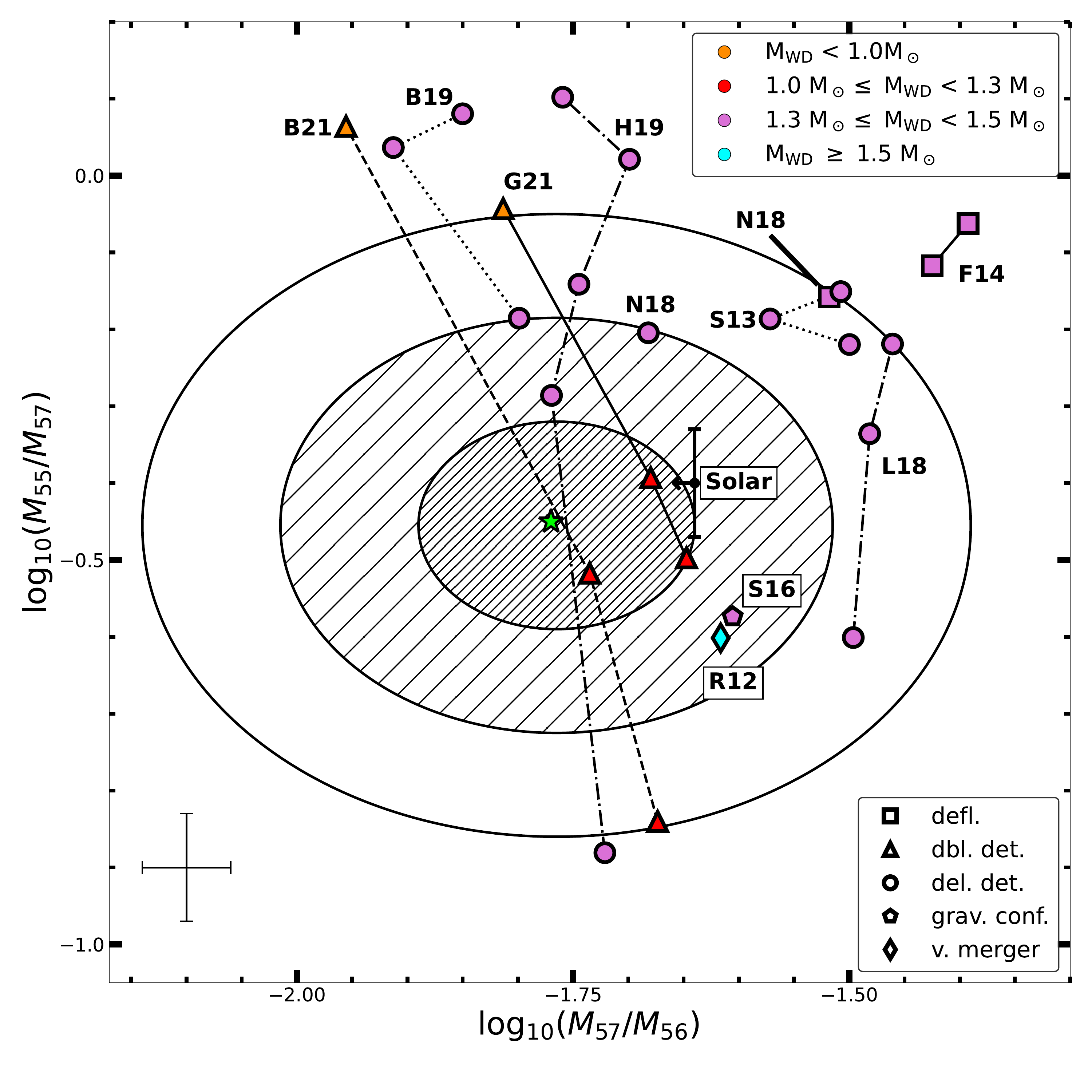}
    \caption{Comparison between our derived isotope ratios for \M{55,56,57} (star and $1,2,3\sigma$ contours) and nucleosynthetic yields for representative explosion models. Models are only shown if they match the \M{56} prior within $3\sigma$ (i.e., $M_{56} \in [0.2,0.8]~M_\odot$). The marker color denotes the mass of the exploding WD and the marker shape indicates the type of explosion model. Models include delayed detonations (del. det., \citealp{ropke2012} (R12), \citealp{seitenzahl2013} (S13), \citealp{leung2018} (L18), \citealp{nomoto2018} (N18), \citealp{hoeflich2019} (H19)), pure deflagrations (defl., \citealp{fink2014} (F14), \citealp{nomoto2018} (N18)), gravitationally-confined detonations (grav. conf., \citealp{seitenzahl2016} (S16)), double detonations (dbl. det., \citealp{gronow2021} (G21), \citealp{boos2021} (B21)), and violent mergers (v. merger, \citealp{ropke2012} (R12)). Line-connected models indicate variations in explosion conditions such as \MWD, \rhoc, or assumptions about the ignition condition(s). Differences in the WD core compositions and/or bulk metallicity are not shown, nor are thick-shell double-detonation models (see \S\ref{sec:discuss}). The Solar value \citep{asplund2009} is potentially an upper limit in this diagram as our isotopic ratios are insensitive to any stable \iFe{56} synthesized by the explosion. The error bar in the lower-left corner represents nucleosynthetic post-processing uncertainties in the models estimated by \citet{bravo2020}. 
    }
    \label{fig:yields}
\end{figure*}

We have presented new \hst photometry of \name extending to $\approx 2400$~days after \tmax and used these observations to construct a pseudo-bolometric light curve. Energy input from \iNi{57} and \iFe{55} are needed to power the late-time luminosity (Fig. \ref{fig:decayfit}). While there are several physical mechanisms which \textit{could} affect the derived isotopic constraints, these scenarios are ultimately disfavored because they produce physically unreasonable results (i.e., $M_{57}\equiv 0 ~M_\odot$ or $M_{55} \equiv 0 ~M_\odot$). We will obtain one additional \hst epoch in Cycle 30 to constrain alternative explanations and improve the isotopic ratio estimates. 

Fig. \ref{fig:yields} compares our derived isotopic ratios to various \snia explosion models produced after the revision of electron-capture rates \citep{dean1998, caurier1999b, caurier1999, brachwitz2000}. We adopt the isotope ratios from the model including time-dependent X-ray escape (see \S\ref{subsec:phot.nuclsyn}) but note that our conclusions are largely independent of the X-ray escape treatment. Some explosion models can be discarded as they do not reproduce \name well near maximum light.\footnote{These comparisons \textit{only} rely on observations of \name and do not account for any discrepancies between models and the ensemble properties of \sneia as a population.} Direct (head-on) WD collision models \citep[e.g., ][]{rosswog2009, vanrossum2016, papish2016} predict isotope ratios inconsistent with observations \citep{papish2016} which is supported by the non-detection of bimodal nebular-phase emission profiles \citep[e.g., ][]{dong2015, vallely2020}. Similarly, graviationally-confined detonations \citep[grav. conf.; e.g., ][]{plewa2004, plewa2007, townsley2007, meakin2009, seitenzahl2016} predict large-scale mixing during the explosion that mix burning by-products into the outer ejecta. This creates strong UV line-blanketing in early spectra, which was not observed for \name. The violent merger model from \citet[][also see \citealp{pakmor2010, pakmor2011,kromer2013}]{ropke2012} can reproduce the near-peak observations of \name and predict isotopic ratios in good agreement with our constraints. However, the violent merger model of \citet{kromer2013} predicts strong [\ion{O}{1}] emission lines due to unburned material near the center of the explosion. This was observed in the nebular spectrum of the subluminous SN Ia~2010lp \citep{taubenberger2013} but [\ion{O}{1}] was not seen in nebular spectra of \name \citep{shappee2013, lundqvist2015, graham2015, taubenberger2015, fransson2015, tucker2022}. Thus, we focus on explosion models consistent with other observational constraints on \name. 

\vspace{0.5cm}

\textbf{Double Detonations} \citep[e.g., ][]{livne1990, livne1991, fink2007, fink2010, woosley2011, gronow2021, boos2021, shen2021} use the ignition of a He-rich surface layer to then drive a compression wave into the WD core that induces the explosion. Bare C/O WD detonations \citep[e.g., ][]{sim2010, shen2018, nomoto2018, bravo2019}, which are qualitatively similar to double-detonation models without the surface ignition, are not shown in Fig. \ref{fig:yields} for clarity but exhibit similar isotopic trends. Double-detonation models initially involved heavy surface shells in order for the He ignition to occur ($M_{\rm{shell}} \gtrsim 0.1~M_\odot$; e.g., \citealp{woosley2011}) and the burning by-products near the surface produced heavy line-blanketing in early-phase optical spectra \citep[e,g,. ][]{polin2019} which is not observed in \name. However, more recent simulations suggest mixing of other elements into the surface shell may facilitate ignition of the He envelope \citep{shen2014, gronow2020} but the minimum mixing and shell mass needed for ignition is still unclear. We exclude double-detonation models with $M_{\rm{shell}} \gtrsim 0.05~M_\odot$ due to the non-detection of He ashes in the early spectra \citep[e.g., ][]{polin2019} and limits on \Ti from the bolometric light curve (see \S\ref{subsec:phot.nuclsyn}), although this cutoff is somewhat arbitrary due to the model uncertainties. Thin-shell double-detonation models involving more massive WD cores ($M_{\rm{WD}} \approx 1.0-1.1~M_\odot$) agree well with our isotopic constraints.

\textbf{Deflagrations} \citep[e.g., ][]{iwamoto1999, travaglio2004, travaglio2005, fink2014, kromer2015, nomoto2018} are subsonic flames that ignite naturally in the core as the WD approaches \Mch due to compressional heating. Fig. \ref{fig:yields} shows these models are broadly inconsistent with our isotopic ratios, likely due to the low amount of high-density burning. However, the density profiles of ``fast'' deflagration models appear to match the density structure of \name reasonably well when modeling early- and nebular-phase spectra \citep{mazzali2014, fransson2015, mazzali2015}. 

\textbf{Delayed Detonations} \citep{ropke2012, seitenzahl2013, ohlmann2014, dessart2014, nomoto2018, leung2018, hoeflich2019, hoeflich2021} start with a deflagration near the center of the star as the WD approaches \Mch but the flame speed transitions from subsonic to supersonic at a specified density \citep{arnett1994, niemeyer1997, khokhlov1997}. The isotopic yields from delayed-detonation models are in poor agreement with our inferred isotopic ratios with the exception of the lowest central density (\rhoc) models of \citet{hoeflich2017, hoeflich2019}. However, the subsonic-to-supersonic transition density ($\rho_{\rm{DDT}}$) is a free parameter which affects the resulting nucleosynthesis. For example, the models of \citet{hoeflich2017, hoeflich2019} show how varying \rhoc with a fixed $\rho_{\rm{DDT}}$ affects the isotopic ratios whereas the models of \citet{bravo2019} assume a single \rhoc and vary $\rho_{\rm{DDT}}$. It remains unclear what areas in the $\rho_c - \rho_{\rm{DDT}}$ parameter space can produce \sneia that match the observed isotopic ratios without introducing new discrepancies with near-peak and nebular-phase observables. 

\vspace{0.5cm}

In summary, all deflagration models are disfavored, as are double-detonation models with low-mass ($M_{\rm{WD}} < 1~M_\odot$) WDs and high-\rhoc delayed-detonation models. The observed isotope ratios fall between the $\rho_c = 2-5\times 10^8~\rm{g}~\rm{cm}^{-3}$ delayed-detonation models from \citet{hoeflich2017, hoeflich2019}, which corresponds to a model-dependent WD mass of $1.2-1.3~M_\odot$ (e.g., Fig. 13 from \citealp{diamond2015}). Furthermore, \citet{mazzali2015} find that the spectra of \name can be reproduced by either a high-energy deflagration explosion \textit{or} a low-energy delayed-detonation explosion, broadly consistent with our isotopic measurements.

Translating the isotope ratios into WD masses is more complicated for double-detonation scenarios. The central pressure ($P_c$) and density (\rhoc) are augmented during the explosion by the converging shockwave produced by the surface ignition \citep{shen2014}. However, the central density and pressure in a degenerate medium are related by $P_c \propto \rho_c^{n}$ with $n=4/3$ ($n=5/3$) for non-relativistic (relativistic) electrons. The double-detonation models in best agreement with our observations have WD masses of $1.0-1.1~M_\odot$ and central densities of $\rho_c = 0.3-1.0\times 10^8~\rm g~\rm{cm}^{-3}$ \citep{boos2021, gronow2021}. Assuming the converging shock increases $P_c$ by a factor of 10, a value well within the predictions of \citet{shen2014}, this leads to an increase in \rhoc of $\sim 5$ which is in rough agreement with the \rhoc estimated from delayed-detonation models. 

However, both classes of explosion models require further study. Multi-dimensional modeling of double-detonation explosions extending to the nebular phase are necessary to evaluate the effect of off-center detonations on the nebular-phase emission line profiles. Conversely, delayed-detonation models require additional simulations to test how these explosions occur without invoking a SD progenitor system since the SD scenario is disfavored for \name (see \S\ref{sec:intro}). DD progenitors producing delayed-detonation explosions are possible \citep[e.g., ][]{piersanti2003} but the available parameter space leading to successful explosions is largely unexplored. Furthermore, WDs with $\rho_c \lesssim 10^9 ~\rm g ~\rm{cm}^{-3}$ are thought to be stable \citep[e.g., ][]{lesaffre2006, wu2019} which poses a challenge to delayed-detonation explosions from low mass ($\lesssim 1.3~M_\odot$) WDs without an external ignition mechanism.

Finally, these results highlight several physical processes in need of additional modeling. It is unclear if the delayed deposition of \iCo{56} positrons (\S\ref{subsec:phot.positrons}, \citealp{kushnir2020}) affects the derived isotope ratios. Clumping in the ejecta has significant ramifications for local energy deposition, such as shortening the recombination time (\S\ref{subsubsec:phot.freezeout}) and decreasing the cooling efficiency of fine-structure emission lines in the MIR (\S\ref{subsec:phot.nuclsyn}). While we disfavor these effects biasing our analyses, stronger predictions will make these scenarios easier to test. Simulations of post-impact companion stars do not agree on the energy dissipation timescale within the stellar envelope (\S\ref{subsubsec:phot.donor}) which makes definitive conclusions about shock-heated companions difficult. Finally, a better understanding of the convective URCA process is required to accurately translate isotopic measurements into constraints on the explosion conditions \citep[e.g., ][]{lesaffre2005, piersanti2022}. These unprecedented observations of \name have provided the first detection of \iFe{55} but also revealed gaps in our knowledge regarding these explosions. 

\vspace{0.5cm}
\section*{Data Availability}

The HST data is publicly available at the MAST archive and our pseudo-bolometric light curve (Fig. \ref{fig:bolcompare}) is included as supplementary material.

\section*{Acknowledgements}

We thank the referee for a very helpful suggestions which improved the manuscript.
We also thank Zach Claytor, Connor Auge, and Michelle Togami for useful discussions. 

M.A.T. acknowledges support from the DOE CSGF through grant DE-SC0019323. CSK and KZS are supported by NSF grants AST-1814440 and AST-1908570. 

Based on observations with the NASA/ESA Hubble Space Telescope obtained from the Data Archive at the Space Telescope Science Institute, which is operated by the Association of Universities for Research in Astronomy, Incorporated, under NASA contract NAS5-26555. Support for program numbers 14678 and 15192 was provided through a grant from the STScI under NASA contract NAS5-26555.

\bibliography{ref}{}
\bibliographystyle{aasjournal}



\end{document}